\documentclass[aps,showpacs,preprint,floatfix]{revtex4}

\usepackage{amsmath}
\usepackage{amssymb}                 
\usepackage{amscd}                   
\usepackage{wasysym}
\usepackage[ansinew]{inputenc}
\usepackage{ae,aecompl}
\usepackage[dvips]{graphicx}
\DeclareGraphicsExtensions{.eps}

\newcommand{\re}{{ \rm e }}

\newcommand{\be}{\begin{equation}}
\newcommand{\ee}{\end{equation}}
\newcommand{\bea}{\begin{eqnarray}}
\newcommand{\eea}{\end{eqnarray}}
\newcommand{\HH}{{\cal H}}
\newcommand{\nn}{\nonumber}

\renewcommand{\vec}[1]{\mathbf{#1}}
\newcommand{\ket}[1]{\left|#1\right\rangle}

\begin{document}
\title{Classical and quantum features of the superfluid to Mott insulator transition}
\author{D. Witthaut, F. Trimborn, and H. J. Korsch}
\affiliation{Fachbereich Physik, TU Kaiserslautern, D--67653 Kaiserslautern, Germany}
\date{\today}

\begin{abstract}
We analyze the correspondence of many-particle and mean-field dynamics for
a Bose-Einstein condensate in an optical lattice.
Representing many-particle quantum states by a classical phase space ensemble
instead of one single mean-field trajectory and taking into account the
quantization of the density by a modified integer Gross-Pitaevskii equation,
it is possible to simulate the superfluid to Mott insulator transition and
other phenomena purely classically.
This approach can be easily extended to higher particle numbers and multidimensional
lattices. Moreover it provides an excellent tool to classify true quantum features
and to analyze the mean-field -- many particle correspondence.

\end{abstract}

\pacs{03.75.Lm, 03.65.Sq}

\maketitle


\textbf{Introduction.}
Ultracold atoms in optical lattices provide an excellent model system for the study
of fundamental condensed matter problems, since experimental parameters can
be controlled in a wide range and with high accuracy. 
If the atomic interactions are weak, bosonic atoms form a Bose-Einstein condensate (BEC).
In the superfluid phase all atoms occupy the same delocalized quantum state and can
be described by a macroscopic wave function. The dynamics of the condensate wave function 
is given in a mean-field approach by the celebrated Gross-Pitaevskii equation (GPE). 
However, mean-field theory takes into account only expectation values and neglects 
quantum fluctuations. Hence it is widely believed to fail when fluctuations are of importance
-- striking examples are number squeezing and quantum phase transitions \cite{Orze01,Grei02}.
In this Letter we propose a phase space ensemble approach based on the mean-field limit
including quantum fluctuations.
If the atomic interactions dominate the dynamics, the system undergoes a quantum phase transition 
from the superfluid (SF) to the Mott insulator (MI) phase even at zero temperature due to quantum 
fluctuations. Since its first experimental observation \cite{Grei02} this
transition attracts more and more attention (see \cite{JaBl05} and references therein).
The MI phase is charaterized by a finite gap in the excitation spectrum and a
vanishing compressibility. Number fluctuations are frozen out and the long-range phase coherence
is lost. In an overall confining potential this leads to the
appearance of a distinct shell structure. This pinning of the on-site occupation number to
integer values has only recently been observed experimentally \cite{Foel06}.

If the optical lattice is sufficiently deep, the dynamics of the atoms is described by 
the Bose-Hubbard model Hamiltonian \cite{Fish89b}
\be
  \hat H = \sum \nolimits_j \left( -J (\hat a_{j+1}^\dagger \hat a_j +
    \hat a_{j}^\dagger \hat a_{j+1}) + \epsilon_j \hat n_j + \frac{U}{2} \hat n_j^2 \right),
    \label{eqn-hami-bh}
\ee
where $\hat a_j$ is the bosonic annihilation operator at the $j$-th lattice site and
$\hat n_j = \hat a_j^\dagger \hat a_j$.
As the Bose-Hubbard model is a genuine many-particle problem, approximations are of exorbitant 
interest. Numerically exact calculations can only be done for very few particles since 
the Hilbert space of the many-particle quantum states grows exponentially both with the 
particle number and the size of the lattice. 

The mean-field approximation for a weakly interacting Bose gas is usually derived within
a Bogoliubov approach, considering the expectation values
$\langle \hat a_j \rangle   = \sqrt{N} \psi_j$ only.
Starting from the Heisenberg equations of motion for the operators $\hat a_j$
and neglecting quantum fluctuations $\langle \hat a_j^\dagger \hat a_j^2  \rangle \approx
\langle \hat a_j^\dagger \rangle \langle \hat a_j  \rangle^2$ yields the discrete GPE 
\be
  i \dot \psi_j = -J (\psi_{j+1} + \psi_{j-1})  + U N |\psi_j|^2 \psi_j
  \label{eqn-dnlse}
\ee
for the classical field $\psi_j$.
This ansatz is exact for vanishing interaction $U=0$ and can be compared to
the Ehrenfest theorem in single particle quantum mechanics.

In this Letter we will propose a phase space approach to the mean-field limit and a 
generalization of the GPE which enables us to describe squeezed states and the SF-MI-transition 
purely classically. In contrast to the established Bogoliubov approach the many-particle state
is represented by an \textit{ensemble} of mean-field trajectories, taking into
account also the higher moments of the many particle quantum state.
The quantization of the density $\hat n_j$ which gives rise to the Mott shell structure is achieved by
the introduction of an integer Gross-Pitaevskii equation (IGPE).
The phase space IGPE dynamics can be easily implemented and efficiently simulated even for
higher particle numbers and two- or three-dimensional lattices.
Moreover we will show that it provides an excellent tool to classify true quantum features
and analyze the mean-field -- many-particle correspondence like the break down of the mean-field
approximation at unstable classical fixed points \cite{CaVa97}.

\textbf{Quantum phase space.}
The key idea of the phase space IGPE approach is to map the quantum many-particle state onto an
\textit{ensemble} of trajectories in classical phase space, whose dynamics
is given by the IGPE. Instead of quantum expectation values we will consider
time-dependent classical ensemble averages.
The initial ensemble is distributed according to a quantum quasi-probability distribution to transfer
the characteristics of the quantum state onto the classical phase space. The Husimi
distribution provides a non-negative and bounded function on quantum phase space, which is hence
particularly suited for a comparison to the classical phase space distribution.
Any many-particle quantum state $\ket{\Psi}$ can be represented by the Husimi
distribution, which is given by the projection onto coherent states
$\ket{ \{ c_j \} }$,
\be
  Q_{\ket{\Psi}}(c_1, \ldots, c_M) = \left| \langle \{ c_j \} |
  \Psi \rangle \right|^2.
   \label{eqn-husimi-lattice}
\ee
In general, the structure of quantum phase space is determined by the operator
algebra generated by the commutation relations of $\hat a_j$ and $\hat a_j^\dagger$.
For the $M$-site Bose-Hubbard model, the relevant algebra is SU($M$),
which reflects the conservation of the number of particles. The generalized coherent
states \cite{Pere86}
\be
  \ket{ \{ c_j \} } = \frac{1}{\sqrt{N!}}
  \bigg( \sum \nolimits_{j=1}^M  c_j \hat a_j^\dagger \bigg)^N \ket{0}
  \label{eqn-coherent}
\ee
are parametrized by the amplitudes at the lattice sites $c_j$ which
are normalized as $\sum \nolimits_j |c_j|^2 = 1$.
These concepts are most easily understood for the special case of just two
lattice sites, where one can introduce the operators $\hat L_x = \frac{1}{2} 
(\hat a_1^\dagger \hat a_2 + \hat a_2^\dagger \hat a_1 ), \; \hat L_y = 
\frac{i}{2} (\hat a_2^\dagger \hat a_1 - \hat a_1^\dagger \hat a_2 )$ and 
$\hat L_z = \frac{1}{2} (\hat a_1^\dagger \hat a_1 - \hat a_2^\dagger \hat a_2) ,$
which form an angular momentum algebra SU(2) with quantum number $\ell = N/2$
\cite{MiSm97}. The Hamiltonian (\ref{eqn-hami-bh}) then can
be rewritten in the Bloch representation as $\hat H =  -2 J \hat L_x + U \hat L_z^2$
up to a constant term. Since the SU(2)-algebra is topological equivalent
to the (Bloch-)sphere $S^2(\theta,\phi)$, the coherent states and the Husimi
distribution can be parametrized by the polar angle $\theta$ and the azimuth angle
$\phi$ via the identification $c_1 = \sqrt{ (1+ \cos \theta)/2 }$
and $c_2 = \re^{i \phi} \sqrt{ (1 - \cos \theta)/2 }$.
The dynamics of the Husimi distribution $Q$ is then given by \cite{Trim07}
\begin{eqnarray}
  \dot Q &=& UN \left( \cos(\theta) - \frac{1}{N} \sin(\theta) \frac{\partial}{\partial \theta}
       \right) \frac{\partial}{\partial \phi} Q  \nn \\
      && -J \left( \sin(\phi) \frac{\partial}{\partial \theta}
      + \cos(\phi) \frac{\cos \theta}{\sin \theta} \frac{\partial}{\partial \phi} \right) Q.
  \label{eqn-husimi-eom}
\end{eqnarray}

\textbf{The integer GPE.}
The classical Hamiltonian function $\HH$ of the mean-field dynamics,
$i d \psi_j/dt = \partial \HH /\partial \psi_j^* $, is given by the expectation
value of the many-particle Hamiltonian (\ref{eqn-hami-bh}), again neglecting 
quantum fluctuations:
\be
  \HH = \sum \nolimits_j \left( -J  (\psi_{j+1}^* \psi_j + c.c.)
    + \epsilon_j |\psi_j|^2 + \frac{UN}{2} |\psi_j|^4 \right).
   \label{eqn-ham-clas}
\ee
However, the quantum fluctuations, in particular the variance of the density, are
of fundamental importance in the MI phase. In principle these fluctuations act like
an effective macroscopic variable driving the system towards integer filling.
This can be already understood in the case of two wells:
One can show that in the limit of strong number squeezing the quantum term 
$\sim \partial^2 Q / \partial \theta \partial \phi$
 in Eq. (\ref{eqn-husimi-eom}) simplifies and the dynamics becomes 
exactly Liouvillian again \cite{Trim07}, however with a modified Hamiltonian 
function including a term describing quantum pressure.
Generalizing this exact result to an extended lattice we propose a modification of 
the GPE including the additional energy
$\Delta \HH = U/2 \sum \nolimits_j (\langle \hat n_j^2 \rangle - \langle n_j \rangle^2)$ 
due to the quantization of the number operator.
This term is only relevant if the interaction energy dominates the
dynamics. In this case the lattice wells decouple and the many-particle wave
function is given by the Gutzwiller ansatz as a direct product
$\ket{\Psi} = \ket{\Psi_1} \otimes \cdots \otimes \ket{\Psi_M}$,
where the local wave functions are written in a Fock basis
$\ket{\Psi_j} = \sum \nolimits_{n_j} b_{n_j} \ket{n_j}$.
Assuming that the local density is given by the mean-field expectation
value $N |\psi_j|^2$, the minimal ansatz for the local quantum state reproducing
this density is given by
\be
  \ket{\Psi_j} = \sqrt{1-d_j} \ket{m_j} + \sqrt{d_j}e^{i\beta} \ket{m_j+1},
\ee
where $m_j$ is the integer part of the local density, i.e.
$N|\psi_j|^2 = m_j + d_j$ with $m_j \in \mathbb{N}$. The variance
of the density is then given by
$ \langle \hat n_j^2 \rangle - \langle n_j \rangle^2 = d_j - d_j^2$. 
This ansatz presumes the smallest possible variance
and thus the minimal energy correction $\Delta \HH$  with respect to the GPE
Hamiltonian (\ref{eqn-ham-clas}).
Including this correction, Hamilton's equations yield an
\textit{integer} Gross-Pitaevskii equation (IGPE)
\be
  i \dot \psi_j = -J (\psi_{j+1} + \psi_{j-1}) + \epsilon_j \psi_j + U (m_j + 1/2) \psi_j.
\ee

\begin{figure}[t]
\centering
\includegraphics[width=6.5cm,  angle=0]{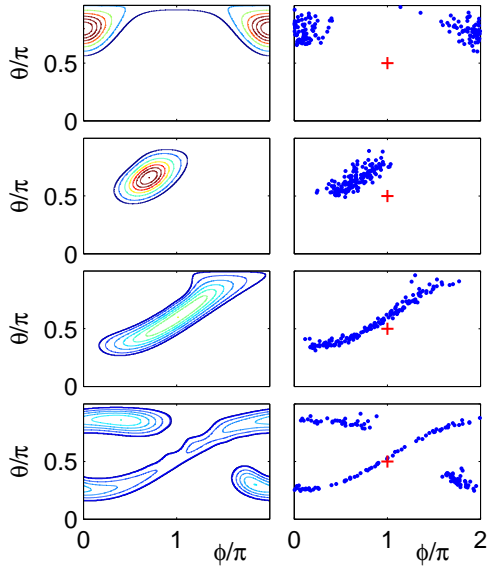}
\caption{\label{fig-den-hyp}
(Color online) Comparison of quantum and classical dynamics in phase
space approaching the hyperbolic fixed point for $UN = 10$, $J=1$
and $N=20$ particles. Left: Evolution of the Husimi density.
Right: Dynamics of an ensemble of 150 classical trajectories
at times $t=0,0.3,0.6,0.9$.}
\end{figure}

\textbf{Bogoliubov vs.~phase space approach.}
As a first illustrative example, we consider the dynamics of the two-mode system.
In the mean-field limit, the quantum expectation value $\langle \hat{\vec L} \rangle/N$
is approximated by the Bloch vector
\be
  \vec s = \frac{1}{2} \left(\begin{array}{c}
  \psi_1^* \psi_2 + \psi_2^* \psi_1 \\
  i (\psi_2^* \psi_1 - \psi_1^* \psi_2) \\
  |\psi_1|^2 - |\psi_2|^2
  \end{array} \right).
\ee
The dynamics of $\vec s$ is restricted to the (Bloch-)sphere, reflecting
the conservation of normalization.
In the non-interacting case $UN = 0$ the system performs common Rabi
oscillations between the two modes.
The anti-symmetric fixed point bifurcates for $UN > 2J$, forming
two novel elliptic and one hyperbolic fixed point. The two novel
fixed points have a non-vanishing population difference $s_z \neq 0$,
leading to the `self-trapping' effect \cite{MiSm97}.

\begin{figure}[t]
\centering
\includegraphics[width=5.5cm,  angle=0]{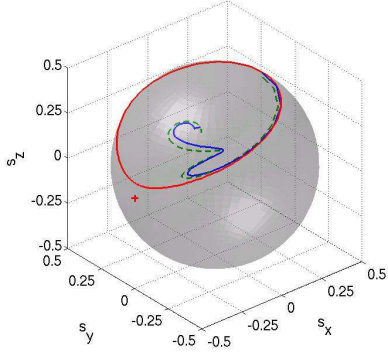}
\caption{\label{fig-hyp-expvalue}
(Color online) Comparison of the many-particle expectation value
$\langle \vec L (t) \rangle/N$ (blue line), the Bogoliubov
approximation $\vec s(t)$ (red line) and the ensemble average
$\langle \vec s (t) \rangle_{\rm cl}$ for 1000 classical trajectories
(dashed green line). Parameters are chosen as in
Fig.~\ref{fig-den-hyp}. }
\end{figure}

At the hyperbolic fixed point the classical dynamics becomes
unstable and the Bogoliubov approach breaks down \cite{CaVa97}.
This is mirrored in quantum phase space as shown in Fig.~\ref{fig-den-hyp}:
The initial coherent state  at $\theta = 0.8 \pi$ and $\phi = 0$ diffuses
in the direction of the unstable manifold.
Therefore the quantum expectation value $\langle \hat{\vec L}(t) \rangle/N$
penetrates into the Bloch sphere in the vicinity of the hyperbolic fixed
point (marked by $+$ in Fig.~\ref{fig-hyp-expvalue}), whereas the single
classical trajectory $\vec s(t)$ stays on the Bloch sphere.
However, this is by no means a breakdown of the mean-field
approximation in phase space. As demonstrated in Fig.~\ref{fig-den-hyp}
the dynamics of a classical {\it ensemble} closely follows the dynamics
of the Husimi distribution. The dynamics of the quantum expectation value
$\langle \hat{\vec L} \rangle/N$ is well reproduced by the ensemble average
$\langle \vec s \rangle_{\rm cl}$ as shown in Fig.~\ref{fig-hyp-expvalue}.

The phase space approach can describe certain features of the
quantum dynamics, such as squeezing and the dynamics at the unstable
classical fixed point. Other aspects of the dynamics are genuine
quantum like interference and tunneling in phase space.
On the opposite, the GPE exactly reproduces the
dynamics of the expectation value $\langle \hat {\vec L} \rangle$
for $U=0$. However, the system is not necessarily `classical'
(cf. \cite{Ball94}).

\begin{figure}[t]
\centering
\includegraphics[width=8cm,angle=0]{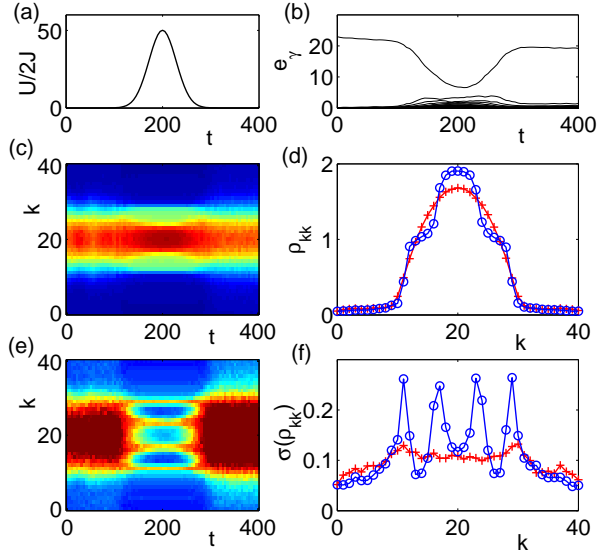}
\caption{\label{fig-mott44time}
(Color online) Dynamics of the SF-MI transition in a harmonic trap with
$\omega = 0.17$, $M=40$ and $N=26$ particles calculated from classical ensemble
averages:
Variation of the parameters (a), eigenvalues of the classical SPDM (b),
density $\rho_{kk}$ (c-d) and its standard deviation $\sigma(\rho_{kk})$(e-f).
Figures (d) and (f) show $\rho_{kk}$ and $\sigma(\rho_{kk})$ in the Mott
phase at time $t=t_c$ calculated with the IGPE (blue circles) and the usual
GPE (red crosses).}
\end{figure}

\begin{figure}[t]
\centering
\includegraphics[width=8.5cm,  angle=0]{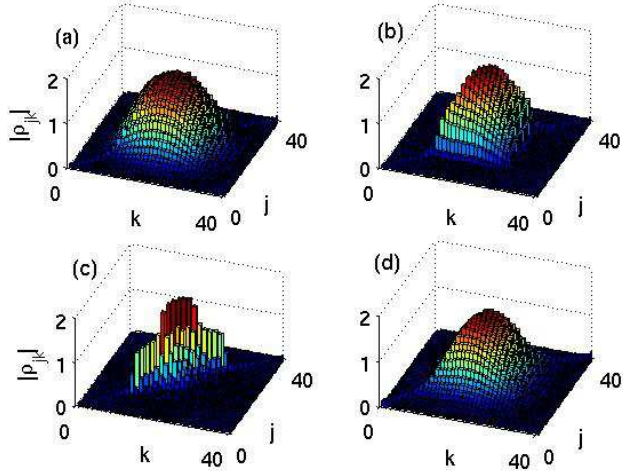}
\caption{\label{fig-mott44den}
(Color online) Dynamics of the SF-MI transition in a harmonic trap:
Magnitude of the classical SPDM $|\rho_{kl}|$ at $t = 0,130,200,400$
(a-d).}
\end{figure}

\textbf{The superfluid to Mott insulator transition.}
The SF-MI transition is considered to be not explicable within mean-field
theory because it is driven by quantum fluctuations. However, the phase space IGPE
method takes into account fluctuations as well as the quantization of the
density $\hat n_j$. Therefore it provides a fully classical description of
the the SF-MI transition.
We consider the dynamics of $N=26$ particles in an optical lattice with a
superimposed harmonic trap described by the Bose-Hubbard Hamiltonian
(\ref{eqn-hami-bh}) with $\epsilon_k = \omega^2 (k - k_0)^2/2$.
We assume that the lattice is initially loaded with a pure BEC
in the Gross-Pitaevskii ground state $\psi_j$. The depth of the lattice is then
increased adiabatically, suppressing the tunneling between the lattice sites,
and decreased back again. This is described by a time-dependent hopping matrix
element $J(t) = J_0[1 + A \re^{-(t-t_c)^2/t_s^2}]^{-1}$ with
$J_0 = 20$, $A=10^3$, $t_c = 200$ and $t_s = 40$, while the interaction
strength $U=1$ is kept constant.
We calculate the time evolution of an ensemble of 200 mean-field trajectories,
whose initial amplitudes $c_k(t=0)$ are distributed according to the Husimi
function (\ref{eqn-husimi-lattice}). A classical approximation of the
single-particle density matrix (SPDM) 
$\rho_{kl}= \langle \hat a_k^\dagger \hat a_l \rangle$
is then provided by the ensemble average
$\rho_{kl}= \langle c_{k}^* c_{l}\rangle_{\rm cl}$.
Figure \ref{fig-mott44time} shows some characteristic features of the MI-SF
transition calculated with the phase space IGPE, reproducing the results
obtained by full many-body calculations \cite{Clar04,Koll04}.
Indeed one observes the occurrence of Mott shells with integer filling,
$\rho_{kk} \in \mathbb{N}$, and small superfluid regions in between.
The density fluctuations are strongly suppressed in the MI phase, whereas
they are significantly stronger between the Mott shells.
In the SF phase one eigenvalue of the SPDM is close to the particle number,
indicating that the many-particle state is a coherent state described by one
single condensate wave function $\psi_j$. In contrast, the MI phase is
characterized by many eigenvalues of the same magnitude. This state cannot
be described by a single condensate wave function, but by a phase space
distribution.
Figure \ref{fig-mott44den} shows the magnitude of the SPDM itself. One
clearly sees that the coherences are lost in the MI phase at $t=t_c$ and
mostly restored at $t=2 t_c$.
The SF-MI transition is reversible to a large extent, which
is demonstrated by increasing $J(t)$ again for $t>t_c$.

Let us notice that the suppression of density fluctuations and the loss and
revival of coherences is already introduced into mean-field theory by the phase
space ensemble approach. However, the usual GPE cannot reproduce the Mott shells and
the energy gap of the particle-hole excitations, because the quantization of the density
is neglected. This becomes most obvious in Fig.~\ref{fig-mott44time} (d) and (f).
The GPE predicts a smooth Thomas-Fermi density profile and a uniform suppression of
density fluctuations in the MI phase.

The occurrence of a gap in the excitation spectrum in the MI phase can also
be understood within the classical phase space approach. Low-energy phonons
cannot be excited as such collective excitations are impossible if the spatial
coherences are lost. The remaining excitations in the MI phase are density
fluctuations, which are discrete in the IGPE description.

\textbf{Conclusion and outlook.}
Summarizing, we map the many particle quantum state to a phase space ensemble obeying a
modified GPE including the backreaction of the quantum fluctuations onto the order 
parameter. The example of the SF-MI phase transition proves the power of the phase space
ensemble method, providing an enormous alleviation of numerical effort and
an illustrative insight into the many-particle dynamics.
For example, the depletion of a BEC at a classically unstable point
\cite{CaVa97} can be fully understood using phase space ensembles.
The atoms are not lost -- they are just redistributed to other modes.
The phase space interpretation gives rise to fundamental questions:
Which properties of a BEC are essentially quantum?
Further work will be devoted to extended three-dimensional
lattices and to a deeper analysis of the classical limit of many-particle
quantum dynamics.
Which criteria determine whether a many-particle system can be simulated
classically? Advanced algorithms for many-particle simulations exploit the
weak entanglement in 1D quantum systems \cite{Vida04} and not the classicality
of the state.

Support from the Studienstiftung des deutschen Volkes and the DFG (GRK 792) is
gratefully acknowledged. We thank M.~Fleischhauer and J.~R.~Anglin for stimulating
discussions.



\end{document}